# Stochastic resolution of identity to CC2 for large systems: excited state properties

Chongxiao Zhao,[1,2,3] Qi Ou,[4] Joonho Lee,[5,*] and Wenjie Dou[2,3,*]

[1]*Department of Chemistry, Zhejiang University, Hangzhou 310027, China*

[2]*Department of Chemistry, School of Science, Westlake University, Hangzhou, Zhejiang 310024, China*

[3]*Institute of Natural Sciences, Westlake Institute for Advanced Study, Hangzhou, Zhejiang 310024, China*

[4]*AI for Science Institute, Beijing 100080, China*

[5]*Department of Chemistry and Chemical Biology, Harvard University, Cambridge, MA 02138, USA*

We apply a stochastic resolution of identity approximation (sRI) to the CC2 method for excitation energy calculations. A set of stochastic orbitals are employed to decouple the crucial 4-index electron repulsion integrals and optimize the contraction steps in CC2 response theory. The CC2 response for excitations builds upon sRI-CC2 ground-state calculations, which scales as $O(N^3)$, where $N$ is a measure for the system size. Overall, the current algorithm for excited states also allows a sharp scaling reduction from original $O(N^5)$ to $O(N^3)$. We test the sRI-CC2 for different molecular systems and basis sets, and we show our sRI-CC2 method can accurately reproduce the results of deterministic CC2 approach. Our sRI-CC2 exhibits an experimental scaling of $O(N^{2.88})$ for a hydrogen dimer chain, allowing us to calculate systems with nearly thousands of electrons.

## 1. INTRODUCTION

Calculating excited state properties for complex molecular systems and extended systems is one of the most challenging tasks in theoretical chemistry. TDDFT is probably one of the most useful means to obtain the excitation energy, due to its affordable computation cost. However, TDDFT often overestimates the energy gaps due to lack of exiton interactions. CC theory is more accurate in predicting excited state properties, however, the high computional cost of the CC method prevents its application to large systems. In all post-HF methods, the implementation of 4-index electron repulsion integrals (ERI) is one of the most challenging steps in electronic structure approaches. High demands for both disk space and computational time hamper their applications to larger systems with hundreds of atoms and have become an intractable bottleneck, e.g., for the coupled cluster (CC) theory.

Among a hierarchy of CC models, CCS, CC2, CCSD, CC3, CCSDT, [1-4] etc., the CC2 model is the most affordable one, which offers accurate ground state energies as well as excitation energies, both correct to second order in the electron fluctuation potential for single-excitation dominated transitions.[5] The CC2 model was first formulated and implemented by Christiansen and Koch *et al*.[6] as an approximation to full CCSD. Compared with conventional standard CCSD model, the CC2 model is capable of the calculation of excitation energies of modest accuracy ($\approx 0.3$ eV)[7] within the framework of CC response theory[8-11] and formally scales as $O(N^5)$ (with N being a measure of the system size).

Over the past 30 years, many attempts to reduce the computational cost associated with 4-index ERIs have been reported. One strategy is to introduce the resolution of identity (RI) approximation[12-17]. A set of pre-optimized auxiliary basis functions are utilized to expensive tensor contractions without sacrificing the accuracy. Another approach is Cholesky decomposition (CD)[18-21], which rearranges the 2-electron ERIs into an iterative process controlled by a single parameter. In both approaches, high rank integrals are disassembled into lower ones, such that storage requirements and CPU time are significantly reduced. However, the overall computational scaling of CC2 remains unaltered as $O(N^5)$, which inevitably hinders application of CC2 for extended systems.

Recently, a stochastic approach to RI approximation, abbreviated as sRI approximation, serves an alternative to handle the costly 4-index ERIs. In the sRI approach, an additional set of random orbitals is employed to further decouple the 4-index ERI tensors. Remarkably, the number of the stochastic orbitals do not need to increase with the system size for intensive quantities, which allows for significant reduction of the computational scalings. Till now, the stochastic orbital appaoaches have been implemented in many quantum chemistry methods, including MP2[22-24], GF2[25-29] and DFT[30,31] *et al*.[32,33]

In a previous work, we have implemented sRI approach to CC2 for ground state calculations[34], where we demonstrate that sRI-CC2 allows for a remarkable scaling reduction from $O(N^5)$ to $O(N^3)$ without significantly affecting the accuracy. In this work, we further extend the sRI approach to CC2 response theory for excitation energy calculations. As tested on a variety of molecular system, we show that our sRI-CC2 response theory not only provides accurate results for excited state as compared against RI-CC2, but also exhibits an overall computational scaling of $O(N^3)$. We expect that the sRI-CC2 response theory will be very useful in describing excited state properties for large systems with 1000 electrons or more.

The manuscript is organized as follows: In the section II, we present the sRI-CC2 theory and the algorithm to calculate excitation energies within the framework of response theory in detail. Section III, we demonstrate the performance of the sRI-CC2 for a variety of molecular systems. We benchmark the sRI-CC2 against results from RI-CC2 in the Q-Chem package[35]. In the section IV, we conclude.

## 2. THEORY

We use the conventional notations in Table 1 to represent the items in the following sections. Furthermore, the total number of atomic orbital (AO) basis functions, auxiliary basis functions, occupied molecular orbitals (MOs), and virtual MOs are denoted by $N_{AO}$, $N_{aux}$, $N_{occ}$, and $N_{virt}$ respectively.

Table 1. Summary of notations in the following equations.

| item | function or indice |
| --- | --- |
| AO Gaussian basis functions | $\chi_\alpha(r_1), \chi_\beta(r_1), \chi_\gamma(r_1), \chi_\delta(r_1), \ldots$ |
| auxiliary basis functions | $P, Q, R, S, \ldots$ |
| general sets of AOs | $\alpha, \beta, \gamma, \delta, \ldots$ |
| general sets of MOs | $p, q, r, s, \ldots$ |
| occupied (active) MOs | $i, j, k, l, \ldots$ |
| unoccupied (virtual) MOs | $a, b, c, d, \ldots$ |

### 2.1. CC2 Excitation Energy Calculation

In CC2 model, the amplitudes equations for singles and doubles can be written as

$$\Omega_{\mu_1} = \langle \mu_1 | \hat{H} + [\hat{H}, T_2] | HF \rangle = 0 \tag{1}$$

$$\Omega_{\mu_2} = \langle \mu_2 | \hat{H} + [F, T_2] | HF \rangle = 0 \tag{2}$$

where $\Omega_{\mu_1}$ and $\Omega_{\mu_2}$ are single and double excitation vector functions. $\mu_1$ and $\mu_2$ are single and double excitation manifolds. $T_2$ is the double excitation cluster operator. |HF⟩ and $F$ are SCF Hartree-Fock reference state and SCF Fock

operator. The Hamiltonian $H$ undergoes a $T_1$-transformation with the exponential function of the single excitation cluster operator $T_1$:

$$\hat{H} = exp(-T_1) H exp(T_1) \tag{3}$$

Similarly, the $T_1$-transformed two-electron MO integrals are given by:

$$(pq\hat{|}rs) = \sum_{\alpha\beta\gamma\delta} \Lambda^p_{\alpha p}\Lambda^h_{\beta q}\Lambda^p_{\gamma r}\Lambda^h_{\delta s}(\alpha\beta|\gamma\delta) \tag{4}$$

$$\Lambda^p = C(I - t_1^T) \tag{5}$$

$$\Lambda^h = C(I + t_1) \tag{6}$$

Here $\Lambda^p$ and $\Lambda^h$ are transformation matrices particle and hole operators. $t_1$ is the auxiliary matrix comprised of singles cluster amplitudes $\{t_i^a\}$:

$$t_1 = \begin{pmatrix} 0 & 0 \\ \{t_i^a\} & 0 \end{pmatrix} \tag{7}$$

The double-excitation amplitudes $\{t_{ij}^{ab}\}$ with their similarity transformed forms $\{\hat{t}_{ij}^{ab}\}$ can be calculated explicitly with a given set of $\{t_i^a\}$:

$$t_{ij}^{ab} = \frac{(ai\hat{|}bj)}{\varepsilon_i - \varepsilon_a + \varepsilon_j - \varepsilon_b} \tag{8}$$

$$\hat{t}_{ij}^{ab} = (1 + \delta_{ij}\delta_{ab})(2 t_{ij}^{ab} - t_{ij}^{ba}) \tag{9}$$

In the RI-CC2 implementation by C. Hättig and F. Weigend[36], explicit expressions for the $\Omega_{\mu_1}$ and $\Omega_{\mu_2}$ have been presented. In our previous work on CC2 ground state properties[34], we have employed the sRI approximation to into these two excitation vector functions, such that we can achieve $O(N^3)$ scaling with iterative solution to the single excitation amplitudes. To further obtain the excitation energies, we need to solve the eigen problem of the Jacobian matrix $A_{\mu_i\nu_j}$ in CC response theory, which is defined as the derivative of the vector function:

$$A_{\mu_i\nu_j} = \frac{\partial\Omega_{\mu_i}}{\partial t_{\nu_j}} = \begin{pmatrix} \langle\mu_1|[\hat{H},\tau_{\nu_1}] + [[\hat{H},\tau_{\nu_1}],T_2]|HF\rangle & \langle\mu_1|[\hat{H},\tau_{\nu_2}]|HF\rangle \\ \langle\mu_2|[\hat{H},\tau_{\nu_1}]|HF\rangle & \delta_{\mu_2\nu_2}\epsilon_{\mu_2} \end{pmatrix} \tag{10}$$

Here the $A_{\mu_2\nu_2}$ block in this Jacobian matrix is diagonal with the matrix elements:

$$\epsilon_{\mu_2} = \epsilon_{aibj} = \epsilon_a - \epsilon_i + \epsilon_b - \epsilon_j \tag{11}$$

The eigen problem of the Jacobian matrix can be solved interatively as well, after we partition the CC2 eigenvalue problem into the following two equations:

$$\sigma_{\mu_1}(\omega, E_{\mu_1}) = \left[ A_{\mu_1 v_1} - \frac{A_{\mu_1 \gamma_2} A_{\gamma_2 v_1}}{\epsilon_{\gamma_2} - \omega} \right] E_{v_1} = A^{eff}_{\mu_1 v_1}(\omega) \, E_{v_1} = \omega E_{\mu_1} \tag{12}$$

$$E_{\mu_2} = -\frac{A_{\mu_2 v_1} E_{v_1}}{\epsilon_{\mu_2} - \omega} \tag{13}$$

Eq. (13) shows that the eigenvector in the double excitation manifold $E_{\mu_2}$ can be obtained from the single one $E_{v_1}$. Such that we can solve the non-linear eigenvalue problem in Eq. (12), which only contains the single excitation manifold.

An initial guess for eigenvalue $\omega$ and for eigenvector $b_{\mu_1}$ can be obtained from models with lower scaling such as CIS or CCS. Similar to $\Omega_{\mu_1}$, the transformed vector $\sigma_{\mu_1}(\omega, b_1)$ is splitted into many terms:

$$\sigma_{ai}(\omega, b_1) = \sigma^0_{ai} + \sigma^G_{ai} + \sigma^H_{ai} + \sigma^I_{ai} + \sigma^J_{ai} \tag{14}$$

$$\sigma^0_{ai} = \sum_b E^{(1)}_{ab} b_{bi} - \sum_j b_{aj} E^{(2)}_{ji} \tag{15}$$

$$\sigma^G_{ai} = +\sum_{dlc} \hat{b}^{cd}_{il} (ld|ac) \qquad \sigma^H_{ai} = -\sum_{dlk} \hat{b}^{ad}_{kl} (ld|ki)$$

$$\sigma^I_{ai} = \sum_{dl} (\hat{b}^{ad}_{il} \hat{F}_{ld} + \hat{t}^{ad}_{il} \bar{F}_{ld}) \qquad \sigma^J_{ai} = \bar{F}'_{ai} \tag{16}\sim(19)$$

$$E^{(1)}_{ab} = \hat{F}_{ab} - \sum_{dlk} \hat{t}^{ad}_{kl} (ld|kb) \qquad E^{(2)}_{ij} = \hat{F}_{ij} + \sum_{dlc} \hat{t}^{cd}_{il} (ld|jc) \tag{20}\sim(21)$$

$$\bar{F}_{ld} = \sum_{ck} [2(ld|kc) - (lc|kd)] b_{ck} \qquad \bar{F}'_{ai} = \sum_{ck} [2(ai|kc) - (ac|ki)] b_{ck} \tag{22}\sim(23)$$

where $\hat{F}$ denotes the similarity transformed Fock matrix

$$\hat{F}_{pq} = \sum_{pq} \Lambda^p_{\mu p} \Lambda^h_{vq} F_{\mu v} + \sum_{ck} [2(pq|ck) - (pk|cq)] t_{ck} \tag{24}$$

The auxiliary vector $\hat{b}^{ab}_{ij}$ in the above equation is defined as

$$\hat{b}^{ab}_{ij} = \frac{2(ai|bj) - (bi|aj)}{\varepsilon_i - \varepsilon_a + \varepsilon_j - \varepsilon_b + \omega} \tag{25}$$

Here we have further defined the modified 4-index ERIs $(ai|bj)$ as

$$(ai|bj) = \hat{P}^{ab}_{ij} \sum_{\alpha\beta\gamma\delta} (\bar{\Lambda}^p_{\alpha a} \Lambda^h_{\beta i} + \Lambda^p_{\alpha a} \bar{\Lambda}^h_{\beta i}) \Lambda^p_{\gamma b} \Lambda^h_{\delta j} (\alpha\beta|\gamma\delta) \tag{26}$$

$$\bar{\Lambda}^p = -C b^T_1 \qquad \bar{\Lambda}^h = C b_1 \tag{27}\sim(28)$$

$$b_1 = \begin{pmatrix} \mathbf{0} & \mathbf{0} \\ \{b^a_i\} & \mathbf{0} \end{pmatrix} \tag{29}$$

$\hat{P}^{ab}_{ij}$ in the above equation is an operator that symmetrizes pairs index $ai$ and $bj$: $\hat{P}^{ab}_{ij} f^{ab}_{ij} = f^{ab}_{ij} + f^{ba}_{ji}$. This concludes the CC2 response theory for excitation energies. The computational cost of the original CC2 and RI-CC2 methods scales as $O(N^5)$. Below, we introduce the algorithm for sRI-CC2 method to achieve scaling of $O(N^3)$.

**2.2. Algorithm for sRI-CC2 Excitation Energy**

With stochastic orbitals, the four index electron integrals are decoupled into product of the stochastic resolution of the identity:

$$(\alpha\beta|\gamma\delta) \approx \frac{1}{N_s}\sum_{\xi=1}^{N_s} R^\xi_{\alpha\beta}R^\xi_{\gamma\delta} \equiv \left\langle R^\xi_{\alpha\beta}R^\xi_{\gamma\delta}\right\rangle_\xi \tag{30}$$

Here $R^\xi_{\alpha\beta}$ is the stochastic resolution of the identity. $\xi$ indicates the number of stochastic orbitals. The sRI approach builds on RI approach. Details of the formulation of RI and sRI approaches can be found in Appendices A and B.

The sRI-CC2 ground state calculation introduced previously is our starting point for the sRI-CC2 response theory. Note that the intermediate quantities such as $\hat{t}^{ab}_{ij}$ and $(ai\bar{|}bj)$ are computed in the ground state calculations. We proceed to compute the transformed vector $\sigma_{\mu_1}(\omega, b_1)$ in CC2 response theory as follow.

1. Calculate the ground-state part with sRI and obtain cluster amplitudes $t_{\mu_1}$ and $\hat{t}_{\mu_2}$. Note that the overall scaling of the sRI-CC2 for ground state calculation is $O(N^3)$.
2. Compute $E^{(1)}_{ab}$ and $E^{(2)}_{ij}$ from Eq. (20)-(21). Initialize the eigenvalue $\omega$ and eigenvector $b_{\mu_1}$.
3. Construct $\overline{\Lambda}^p$ and $\overline{\Lambda}^h$ from Eq. (27)-(28).
4. Respectively obtain each partitions of the transformed vector and eventually add up to $\sigma_{\mu_1}(\omega, b_1)$. Intermediates $(ai\bar{|}bj)$ and $\hat{b}_{\mu_2}$ are utilized "on the fly" with no need to calculate and store. Take the construction of $\sigma^G_{ai}$ as an example

$$(ai\bar{|}bj) = \hat{P}^{ab}_{ij}\sum_{\alpha\beta\gamma\delta}(\overline{\Lambda}^p_{\alpha a}\Lambda^h_{\beta i} + \Lambda^p_{\alpha a}\overline{\Lambda}^h_{\beta i})\Lambda^p_{\gamma b}\Lambda^h_{\delta j}\left\langle R^\xi_{\alpha\beta}R^\xi_{\gamma\delta}\right\rangle_\xi$$

$$= \hat{P}^{ab}_{ij}\left\langle \sum_{\alpha\beta}(\overline{\Lambda}^p_{\alpha a}\Lambda^h_{\beta i} + \Lambda^p_{\alpha a}\overline{\Lambda}^h_{\beta i})R^\xi_{\alpha\beta}\hat{R}^\xi_{bj}\right\rangle_\xi = \hat{P}^{ab}_{ij}\left\langle \overline{R}^\xi_{ai}\hat{R}^\xi_{bj}\right\rangle_\xi = \left\langle \overline{R}^\xi_{ai}\hat{R}^\xi_{bj} + \overline{R}^\xi_{bj}\hat{R}^\xi_{ai}\right\rangle_\xi \tag{31}$$

$$\sigma^G_{\mu_1} = \sum_{abj}\hat{b}^{ab}_{ij}(jb\hat{|}ca)$$

$$= \sum_{abj}\frac{2(ai\bar{|}bj) - (bi\bar{|}aj)}{\varepsilon_i - \varepsilon_a + \varepsilon_j - \varepsilon_b + \omega}\left\langle \hat{R}^{\xi'}_{jb}\hat{R}^{\xi'}_{ca}\right\rangle_{\xi'}$$

$$= \left\langle\sum_{abj}\frac{2(\overline{R}^\xi_{ai}\hat{R}^\xi_{bj} + \overline{R}^\xi_{bj}\hat{R}^\xi_{ai}) - (\overline{R}^\xi_{bi}\hat{R}^\xi_{aj} + \overline{R}^\xi_{aj}\hat{R}^\xi_{bi})}{\varepsilon_i - \varepsilon_a + \varepsilon_j - \varepsilon_b + \omega}\hat{R}^{\xi'}_{jb}\hat{R}^{\xi'}_{ca}\right\rangle_{\xi\xi'} \tag{32}$$

Each item in Eq. (32) can be transformed into the expression of time integrals

$$\frac{\bar{R}^{\xi}_{ai}\hat{R}^{\xi}_{bj}}{\varepsilon_i - \varepsilon_a + \varepsilon_j - \varepsilon_b + \omega}\hat{R}^{\xi'}_{jb}\hat{R}^{\xi'}_{ca} = -\int_0^{\infty}\left[\hat{R}^{\xi}_{bj}e^{(\varepsilon_j-\varepsilon_b)t}\hat{R}^{\xi'}_{jb}\right]\bar{R}^{\xi}_{ai}e^{(\varepsilon_i-\varepsilon_a)t}\hat{R}^{\xi'}_{ca}e^{\omega t}dt \quad (33)$$

$$\frac{\bar{R}^{\xi}_{bi}\hat{R}^{\xi}_{aj}}{\varepsilon_i - \varepsilon_a + \varepsilon_j - \varepsilon_b + \omega}\hat{R}^{\xi'}_{jb}\hat{R}^{\xi'}_{ca} = -\int_0^{\infty}\left[\bar{R}^{\xi}_{bi}e^{(\varepsilon_i+\varepsilon_j-\varepsilon_b)t}\hat{R}^{\xi'}_{jb}\right]\hat{R}^{\xi}_{aj}e^{(-\varepsilon_a)t}\hat{R}^{\xi'}_{ca}e^{\omega t}dt \quad (34)$$

Items in the brackets are calculated first to decouple the indices and these four items cost $O(N_sN_tN^3)$, where $N_s$ is the number of stochastic orbitals and $N_t$ is the number of quadrature points, about 10 with the assistance of Laplace transform. Since both $N_s$ and $N_t$ are small prefactors, the scaling is thus reduced to $O(N^3)$. All sub-items are processed in this way to complete $\sigma_{\mu_1}(\omega, b_1)$.

5. Once the transformed vector $\sigma^{(i)}_{\mu_1} = \sigma_{\mu_1}(\omega^{(i)}, b_1^{(i)})$ is constructed in iteration $i$, an updated eigenvalue $\omega^{(i+1)}$ and the residual are calculated according to the following equations

$$\omega^{(i+1)} = \frac{\sigma^{(i)}_{\mu_1}b^{(i)}_{\mu_1}}{||b^{(i)}_1||^2} \quad (35)$$

$$r^{(i)}_{\mu_1} = \frac{\sigma^{(i)}_{\mu_1} - \omega^{(i+1)}b^{(i)}_{\mu_1}}{||b^{(i)}_1||\,n_{occ}n_{virt}} \quad (36)$$

, where $n_{occ}$ and $n_{virt}$ respectively denote the number of occupied and virtual MOs.

6. If the Frobenius norm of the residual $||r^{(i)}_{\mu_1}||$ is larger than a pre-set threshold, e.g., $10^{-5}$, a perturbational estimate $u^{(i)}_{\mu_1}$ is formed to modify the eigenvector $b^{(i)}_{\mu_1}$, else the iteration is stopped.

$$u^{(i)}_{\mu_1} = \frac{r^{(i)}_{\mu_1}}{\varepsilon_i - \varepsilon_a} \quad (37)$$

$$b^{(i+1)}_{\mu_1} = \frac{b^{(i)}_{\mu_1} + u^{(i)}_{\mu_1}}{||b^{(i)}_1||} \quad (38)$$

and a DIIS algorithm for $b_{\mu_1}$ can accelerate this procedure.

7. Repeat steps 3-6 until self-consistency is reached.

From the steps above, we can see that except steps 1 and 4, scaling as $O(N^3)$, other steps mainly cost $O(N^2)$. In summary, the overall cost of sRI-CC2 excitation energy calculation is theoretically reduced to $O(N^3)$.

## 3. RESULTS AND DISCUSSION

To test the accuracy and efficiency of our sRI-CC2, we apply it to a variety of systems with different basis sets for the lowest singlet excitation energies. All the sRI-CC2 energies or CPU time are averaged over 10 different randomly selected sampling seeds. The standard deviations are estimated by the 10 runs (note that the error in the stardard deviation is $0.746\sigma$ for the 10 finite number of runs). Meanwhile, we utilize the RI-CC2 in the Q-Chem program package with same parameters for comparison.

### 3.1. Hydrogen Dimer Chains $H_n$

We first apply our method to a series of hydrogen dimer chains $H_n$ with sto-3g basis and 1600 stochastic orbitals as test samples, with n being the number of hydrogen atoms ranging from 10 to 300. The distance between two hydrogen atoms in each dimer is 0.74 Å, and the distance between two hydrogen atoms of each adjacent dimer is 1.26 Å.

In Table 2, we list the lowest excitation energies from sRI-CC2 in comparison with RI-CC2 in Q-Chem. We have also list the errors (the difference between the sRI results and the RI results) and standard deviations from 10 runs. We first observe that the CC2 excitation energies slowly decrease with the number of H atoms, in both RI and sRI calculations. Notice also that the results from sRI-CC2 and RI-CC2 agree well with each other, where the error between sRI and RI is always less than 0.2 eV and this would further decrease if we utilize more stochastic orbitals. We note that the standard deviation does not increase with the system size for fixed number of stochastic orbitals. Since the exciation energies are intensive properties, we expect that we do not need to increase the number of stochastic orbitals when increasing the system size.

Table 2. Lowest excitation energies (in eV) for a series of hydrogen dimer chains with sto-3g basis. Here $N_H$ is the number of hydrogen in the corresponding dimer chain, the same value as the number of correlated electrons $N_e$. The standard deviation $\sigma$ is calculated from 10 independent samples and in eV.

| $N_H$ | RI | sRI | error | std deviation |
|---|---|---|---|---|
| $H_{10}$ | 16.7772 | 16.8304 | 0.0532 | 0.3290 |
| $H_{50}$ | 14.9868 | 14.9688 | 0.0180 | 0.3134 |
| $H_{100}$ | 14.8981 | 14.7801 | 0.1180 | 0.1620 |
| $H_{200}$ | 14.8721 | 15.0026 | 0.1305 | 0.2724 |
| $H_{300}$ | 14.8666 | 14.6736 | 0.1930 | 0.3493 |

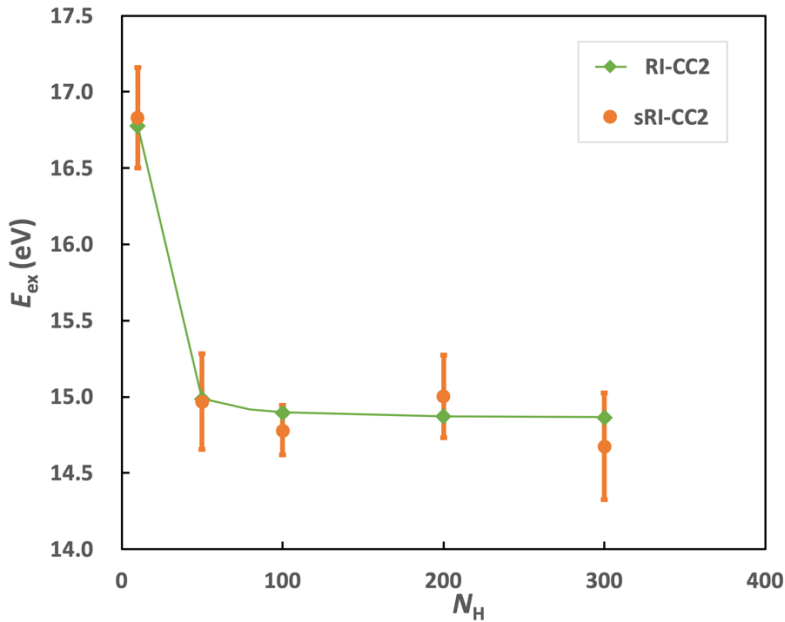

Figure. 1. Excitation energy for sRI-CC2 and RI-CC2 (obtained in Q-Chem) as a function of the number of hydrogen atoms with $N_s$ = 1600. The error bar is from the standard deviation calculated from 10 different seeds.

A more direct comparison is shown in Figure 1. The error bar indicates the standard deviations of 10 independent runs. Results from our sRI-CC2 agree well with RI-CC2 method and all the values from RI-CC2 fall in the error bars, which indicates our sRI-CC2 provides a modest agreement with the standard results. Again, we observe that the statistical error bar does not increase with the system size of hydrogen dimer chains. This observation indicates that we can use fixed number of stochastic orbtials to achieve consistent accuracy for small and large systems. This claim has been made in stochastic MP2[22-24], GF2[25-29], DFT[30,31] and GW[33,37-39] methods as well.

We then test the computational scaling of our sRI-CC2 method. In Figure 2, we plot the CPU time for sRI-CC2 and RI-CC2 calculations as a function of the number of hydrogen atoms. All the calculations are implemented with the high-performance computing (HPC) with Intel CPU, utilizing a single compute node and 64 computational cores. We notice that we need roughly 20 cycles for the SCF interations for convergence in sRI-CC2 calculations, slightly larger than RI-CC2 number of iterations. The overall CPU time for sRI-CC2 is averaged over 10 sampling seeds. Notice that sRI-CC2 exhibits an experimental scaling of $O(N_H^{2.88})$, slightly better than the theoretical scaling of $O(N_H^3)$. RI-CC2 exhibits an experimental scaling of $O(N_H^{4.68})$, which is close to the theoretical scaling of $O(N_H^5)$. The crossover happens at $N_H = 300$, meaning that for larger system, sRI-CC2 will exceed RI-CC2 perfermance. We are currently working on implementation of the sRI-CC2 in QChem package, such that we can further reduce the prefactor of the computational cost in sRI-CC2 calculations and apply the sRI-CC2 method to systems with 1000 electrons or more.

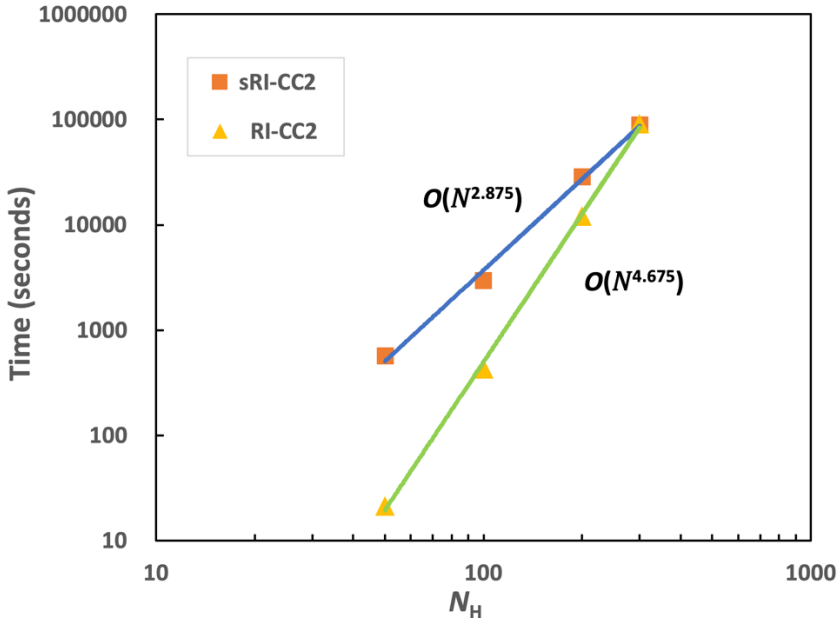

Figure. 2. CPU time as a function of the number of hydrogen atoms with $N_s = 1600$. The computational time of sRI-CC2 is averaged over 10 different runs.

### 3.2. (All-E)-Olefin Chains $C_{2n}H_{2n+2}$

To test our methods on more realistic molecular systems, we further apply sRI-CC2 to a series of (all-E)-alkenes $C_{2n}H_{2n+2}$, with n ranging from 1 to 8. Here, we have used cc-pVDZ and aug-cc-pVDZ as our basis sets. The experimental geometries[39] are provided in the Supporting Information. We have also used 800 stochastic orbitals for sRI calculations.

In Table 3, we list the lowest excitation energies from RI and sRI calculations. The error indicates the difference between RI results and sRI results. The standard deviations are estimated from 10 sRI runs. To better illustrate our results, we further plot the lowest excitation energies as a function of system size in Figure 3. The error bar indicates the standard deviation in sRI runs. We first notice that the excitation energies decrease with the system size, reaching to a plateau eventually. In addition, from the table and the figure, we observe that sRI results agree well with RI results, with all errors lying in the error bar (standard deviation). Notice also that neither the errors and the standard deviation increases with the system size. Finally, we observe that the stochastic error does not strongly depends on the basis set:

when we increase basis from cc-pvdz to aug-cc-pvdz, the stochastic error does not increase noticeably.

Table 3. Results for of (all-E)-$C_{2n}H_{2n+2}$ alkenes with n = 1~5. Here $N_e$ is the number of correlated electrons. The excitation energies from RI and sRI, their errors, and the standard deviations of 10 sRI runs are in eV.

| $N_e$ | molecule | cc-pVDZ | | | | aug-cc-pVDZ | | | |
|---|---|---|---|---|---|---|---|---|---|
| | | RI | sRI | error | std deviation | RI | sRI | error | std deviation |
| 16 | $C_2H_4$ | 8.7502 | 8.5973 | 0.1529 | 0.4688 | 7.1618 | 7.2212 | 0.0594 | 0.1998 |
| 30 | $C_4H_6$ | 6.6371 | 6.6706 | 0.0335 | 0.4164 | 6.1335 | 6.1205 | 0.0130 | 0.2352 |
| 44 | $C_6H_8$ | 5.4894 | 5.6185 | 0.1291 | 0.2619 | 5.5858 | 5.2959 | 0.2899 | 0.3512 |
| 58 | $C_8H_{10}$ | 4.7597 | 4.6807 | 0.0790 | 0.3017 | 5.2470 | 5.1938 | 0.0532 | 0.2745 |
| 72 | $C_{10}H_{12}$ | 4.3393 | 4.3124 | 0.0269 | 0.2963 | 5.0650 | 5.1361 | 0.0711 | 0.3080 |

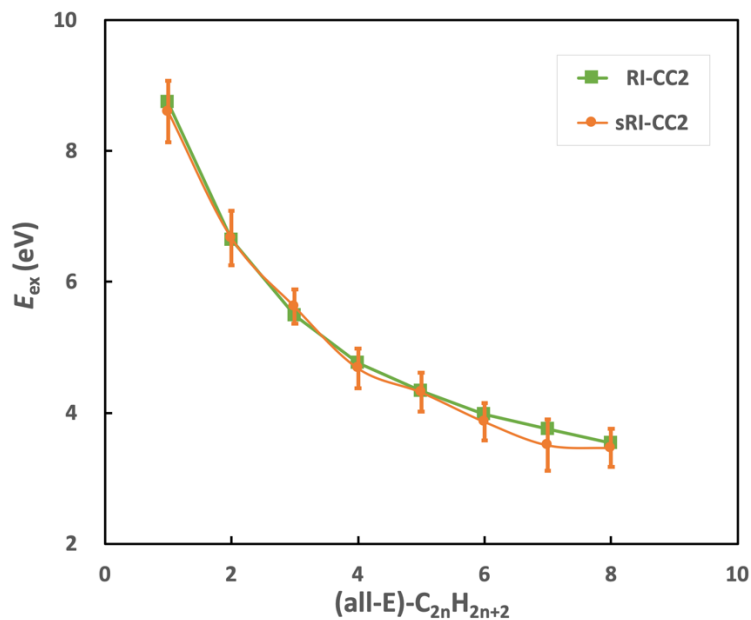

Figure. 3. Excitation energy for a series of (all-E)-$C_{2n}H_{2n+2}$ alkenes with cc-pVDZ basis set and $N_s$ = 800. The error bar is from the standard deviation calculated from 10 different seeds.

In Figure 4, we plot the overall computational time of sRI-CC2 for $C_{2n}H_{2n+2}$ alkenes as a function of system size with different number of stochastic orbitals $N_s$ = 800, 1600 and 3200. Note that regardless the number of stochastic orbitals, the overall scacling of the sRI-CC2 is around $O(N_e^{2.6})$ or $O(N_e^{2.8})$, slightly better than the theoretical scaling $O(N_e^3)$. The RI-CC2 exhibits overall scaling of $O(N_e^{4.31})$, close to the theoretical scaling of $O(N_e^5)$. The crossover happens around $N_e = 60$, where the computational cost of RI-CC2 exceeds sRI-CC2 with 800 stochastic orbitals for larger system size.

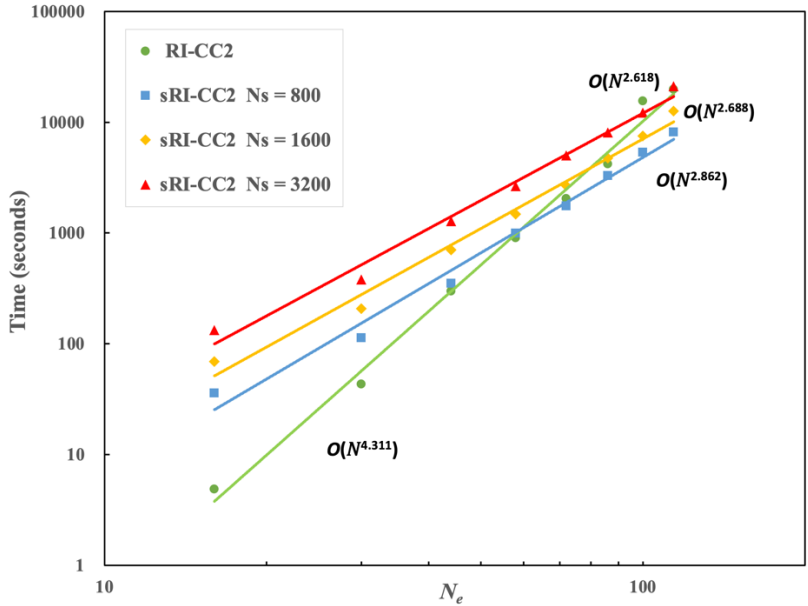

Figure. 4. CPU time as a function of the number of correlated electrons with $N_s$ = 800, 1600 and 3200. The computational time of sRI-CC2 is averaged from 10 different runs.

### 3.3. Molecular Systems

Finally, we test the performance of sRI-CC2 method for a variety of molecules. The experimental geometries[39,40] of the molecules are provided in the Supporting Information. All results from sRI-CC2 are calculated with $N_S$ = 800 stochastic orbtials. cc-pVDZ and aug-cc-pVDZ are chosen as our basis sets. The mean values, errors, and standard deviations are list in Table 4. Overall sRI-CC2 agrees with RI-CC2 well, with all errors lying in the standard deviations. Again, the stochastic error does not increase noticeably with the basis set. This concludes our claims that our sRI-CC2 will be useful for a variety of molecular systems.

Table 4. Results for of 11 different molecules. Here $N_e$ is the number of correlated electrons. The excitation energies from Q-Chem and sRI, their errors, and the standard deviations of 10 sRI runs are in eV.

| $N_e$ | molecule | cc-pVDZ | | | | aug-cc-pVDZ | | | |
|---|---|---|---|---|---|---|---|---|---|
| | | RI | sRI | error | std deviation | RI | sRI | error | std deviation |
| 2 | $H_2$ | 14.0308 | 13.9505 | 0.0803 | 0.4111 | 12.7056 | 12.6157 | 0.0899 | 0.2679 |

| | | | | | | | | |
|---|---|---|---|---|---|---|---|---|
| 10 | $H_2O$ | 8.1101 | 8.1451 | 0.0350 | 0.5343 | 7.1020 | 7.0888 | 0.0132 | 0.4497 |
| 2 | He | 52.5900 | 52.3018 | 0.2882 | 0.9645 | 21.8743 | 21.6330 | 0.2413 | 0.3555 |
| 4 | Be | 5.4347 | 5.2817 | 0.1530 | 0.2755 | 5.2685 | 5.2349 | 0.0336 | 0.2285 |
| 10 | Ne | 50.2279 | 49.8211 | 0.4068 | 0.7243 | 18.7441 | 18.2055 | 0.5386 | 0.5642 |
| 10 | $CH_4$ | 12.3851 | 12.2510 | 0.1341 | 0.4079 | 10.4742 | 10.4117 | 0.0625 | 0.3125 |
| 4 | LiH | 3.7399 | 3.8086 | 0.0687 | 0.4123 | 3.7841 | 3.8484 | 0.0643 | 0.4187 |
| 12 | LiF | 5.1234 | 5.1297 | 0.0063 | 0.4329 | 5.4413 | 5.2285 | 0.2128 | 0.3668 |
| 10 | HF | 10.5782 | 10.4202 | 0.1580 | 0.4481 | 9.8279 | 9.7500 | 0.0779 | 0.4019 |
| 22 | c-$C_3H_4$ | 7.1215 | 7.2194 | 0.0979 | 0.6223 | 6.6596 | 6.5410 | 0.1186 | 0.3419 |
| 40 | $EtCONH_2$ | 5.9118 | 5.9113 | 0.0005 | 0.4992 | 5.7510 | 5.5906 | 0.1604 | 0.5020 |

## 4. CONCLUSIONS

We develop a stochastic implementation of the RI-CC2 method for excitation energy calculations, namely sRI-CC2 theory. With the introduction of stochastic orbitals, the expensive 4-index ERIs are decoupled into lower-rank integrals, allowing for a reduction of computational scaling from the original $O(N^5)$ to $O(N^3)$. This work extends our previous implementation of sRI-CC2 for ground state to excited state, achieving efficient and accurate results for excitation energies within the CC2 model. As tested on a variety of molecular systems, sRI-CC2 provides exitaiton energies agree well with RI-CC2 method, demonstrating wide application of the sRI-CC2 method. We expect that the sRI-CC2 will be very useful in predicting excited states for very large systems.

## ASSOCIATED CONTENT

**Supporting Information**

The Supporting Information is available free of charge at {molecule_geometries.txt}.

Information as mentioned in the text. (TXT)

## AUTHOR INFORMATION

**Corresponding Authors**

*E-mail: joonholee@g.harvard.ed (J.L.).

*E-mail: douwenjie@westlake.edu.cn (W.D.).

**ORCID**

Chongxiao Zhao: 0009-0009-8627-8911

Joonho Lee: 0000-0002-9667-1081

Wenjie Dou: 0000-0001-5410-6183

**Notes**

The authors declare no competing financial interest.

**ACKNOWLEDGEMENT**

We acknowledge funding from National Natural Science Foundation of China and the startup funding from Westlake University. J. L. acknowledges the startup funding from Harvard University. C. Z. thank Jian Zhu for helpful discussion. We also acknowledge high performance computing (HPC) service at Westlake University.

**APPENDIX**

**A. Deterministic Resolution of Identity**

The 4-, 3- and 2-index ERIs are defined as

$$(\alpha\beta|\gamma\delta) = \iint dr_1\, dr_2 \frac{\chi_\alpha(r_1)\chi_\beta(r_1)\chi_\gamma(r_2)\chi_\delta(r_2)}{r_{12}} \tag{39}$$

$$(\alpha\beta|P) = \iint dr_1\, dr_2 \frac{\chi_\alpha(r_1)\,\chi_\beta(r_1)\,\chi_P(r_2)}{r_{12}} \tag{40}$$

$$V_{PQ} = (P|Q) = \iint dr_1\, dr_2 \frac{\chi_P(r_1)\,\chi_Q(r_2)}{r_{12}} \tag{41}$$

The RI approximation enables the 4-index ERIs to be expanded into lower-rank integrals with auxiliary basis $\{P\}$:

$$(\alpha\beta|\gamma\delta) \approx \sum_{PR}^{N_{aux}} (\alpha\beta|P)[V^{-1}]_{PR}(R|\gamma\delta)$$

$$= \sum_Q^{N_{aux}} \left[ \sum_P^{N_{aux}} [(\alpha\beta|P)[V^{-1/2}]_{PQ}] \right] \left[ \sum_R^{N_{aux}} [V^{-1/2}]_{QR}(R|\gamma\delta) \right] \tag{42}$$

Defining

$$K_{\alpha\beta}^{Q} \equiv \sum_{P}^{N_{aux}} (\alpha\beta|P) V_{PQ}^{-1/2} \tag{43}$$

which, scaling as $O(N_{AO}^2 N_{aux}^2)$. Then the expression in Eq. (42) can be simplified as

$$(\alpha\beta|\gamma\delta) \approx \sum_{Q}^{N_{aux}} K_{\alpha\beta}^{Q} K_{\gamma\delta}^{Q} \tag{44}$$

Besides, since ERIs are commonly utilized in MO basis, their transformation from AO basis to MO basis can be completed in two steps, both costing $O(N_{AO}^3 N_{aux})$,

$$K_{p\beta}^{Q} = \sum_{\alpha}^{N_{AO}} C_{\alpha}^{p} K_{\alpha\beta}^{Q} \tag{45}$$

$$K_{pq}^{Q} = \sum_{\beta}^{N_{AO}} C_{\beta}^{q} K_{p\beta}^{Q} \tag{46}$$

where $C_{\alpha}^{p}$ and $C_{\beta}^{q}$ is the usual SCF MO coefficient matrix.

Both $N_{AO}$ and $N_{aux}$ are dependent on the system size. Therefore, according to Eqs. (42)-(46), the overall cost of RI-ERIs scales as $O(N_{AO}^4 N_{aux})$, which we usually label as $O(N^5)$.

## B. Stochastic Resolution of Identity

The stochastic optimization of the RI approximation adopts another set of stochastic orbitals $\{\theta^{\xi}\}$, $\xi = 1, 2, \ldots, N_S$. All these stochastic orbitals are column arrays of length $N_{aux}$ with each random element $\theta_A^{\xi} = \pm 1$. They satisfy the following equations:

$$\theta_A^{\xi} \theta_A^{\xi} = 1 \qquad \theta_A^{\xi} \theta_B^{\xi} = \pm 1 \tag{47}\sim(48)$$

$$\langle \theta \otimes \theta \rangle_{\xi} = \frac{1}{N_S} \sum_{\xi=1}^{N_S} \theta^{\xi} \otimes (\theta^{\xi})^T = \begin{pmatrix} \langle \theta_1 \theta_1 \rangle_{\xi} & \langle \theta_2 \theta_1 \rangle_{\xi} & \cdots & \langle \theta_1 \theta_{N_{aux}} \rangle_{\xi} \\ \langle \theta_2 \theta_1 \rangle_{\xi} & \langle \theta_2 \theta_2 \rangle_{\xi} & & \langle \theta_2 \theta_{N_{aux}} \rangle_{\xi} \\ \vdots & & \ddots & \vdots \\ \langle \theta_{N_{aux}} \theta_1 \rangle_{\xi} & \langle \theta_{N_{aux}} \theta_2 \rangle_{\xi} & \cdots & \langle \theta_{N_{aux}} \theta_{N_{aux}} \rangle_{\xi} \end{pmatrix} \approx I \tag{49}$$

Accordingly, we insert this approximate identity matrix in Eq. (49) into the 4-index ERI and obtain its stochastic form:

$$(\alpha\beta|\gamma\delta) \approx \sum_{QS}^{N_{aux}} \sum_{PR}^{N_{aux}} (\alpha\beta|P) V_{PQ}^{-1/2} I_{QS} V_{SR}^{-1/2} (R|\gamma\delta)$$

$$\begin{aligned}
&= \sum_{QS}^{N_{aux}} \sum_{PR}^{N_{aux}} (\alpha\beta|P) V_{PQ}^{-1/2} \left(\langle \theta \otimes \theta^T \rangle_\xi\right)_{QS} (R|\gamma\delta) V_{SR}^{-1/2} \\
&= \left\langle \left[\sum_P^{N_{aux}} (\alpha\beta|P) \sum_Q^{N_{aux}} \left(V_{PQ}^{-1/2}\theta_Q\right)\right]\left[\sum_R^{N_{aux}} (R|\gamma\delta) \sum_S^{N_{aux}} \left(V_{SR}^{-1/2}\theta_S^T\right)\right]\right\rangle_\xi
\end{aligned} \quad (50)$$

Similarly, we define the ξth element of the stochastic average as:

$$R_{\alpha\beta}^\xi = \sum_P^{N_{aux}} (\alpha\beta|P) \left[\sum_Q^{N_{aux}} \left(V_{PQ}^{-1/2}\theta_Q\right)\right] \equiv \sum_P^{N_{aux}} (\alpha\beta|P) L_P^\xi \quad (51)$$

where the construction of $L_P^\xi$ costs $O(N_s N_{aux}^2)$ and $R_{\alpha\beta}^\xi$ costs $O(N_s N_{AO}^2 N_{aux})$. Thus the 4-index ERI can be rewritten as a stochastic average and scales as $O(N_s N_{AO}^4)$:

$$(\alpha\beta|\gamma\delta) \approx \frac{1}{N_s} \sum_{\xi=1}^{N_s} R_{\alpha\beta}^\xi R_{\gamma\delta}^\xi \equiv \left\langle R_{\alpha\beta}^\xi R_{\gamma\delta}^\xi \right\rangle_\xi \quad (52)$$

The ERI in the AO basis can be transformed to MO basis in the same way as RI, scaling as $O(N_s N_{AO}^3)$,

$$R_{p\beta}^\xi = \sum_\alpha^{N_{AO}} C_\alpha^p R_{\alpha\beta}^\xi \quad (53)$$

$$R_{pq}^\xi = \sum_\beta^{N_{AO}} C_\beta^q R_{p\beta}^\xi \quad (54)$$

The overall calculation of sRI-ERIs approximately scales as $O(N^4)$ for large-size systems since with the expansion of the system size, the $N_s$ is almost unchanged and relatively small compared with $N_{AO}$ and $N_{aux}$. Furthermore, since we only employ 4-index ERIs as intermediates and contract them with 3-index ERIs "on the fly", the actual cost can be further reduced in our sRI-CC2 algorithm.

**DATA AVAILABILITY**

The data that support the findings of this study are available within the article and supplementary material.